\documentclass[aip,reprint,superscriptaddress]{revtex4-1}
\usepackage[dvipdfmx]{graphicx}
\usepackage{amsmath,cases}
\usepackage[colorlinks,linkcolor=blue,urlcolor=blue]{hyperref}
\usepackage{siunitx}
\begin{document}

\title{Switching of magnon parametric oscillation by magnetic field direction}

\author{Sohei Horibe}
\email{horibe-sohei035@g.ecc.u-tokyo.ac.jp}
\affiliation{Department of Applied Physics, Faculty of Engineering, University of Tokyo, Tokyo 113-8656, Japan}

\author{Hiroki Shimizu}
\affiliation{Department of Applied Physics, Faculty of Engineering, University of Tokyo, Tokyo 113-8656, Japan}

\author{Koujiro Hoshi}
\affiliation{Department of Applied Physics, Faculty of Engineering, University of Tokyo, Tokyo 113-8656, Japan}
\affiliation{Institute for AI and Beyond, The University of Tokyo, Tokyo 113-8656, Japan}

\author{Takahiko Makiuchi}
\affiliation{Department of Applied Physics, Faculty of Engineering, University of Tokyo, Tokyo 113-8656, Japan}
\affiliation{Quantum-Phase Electronics Center, University of Tokyo, Tokyo 113-8656, Japan}

\author{Tomosato Hioki}
\affiliation{Department of Applied Physics, Faculty of Engineering, University of Tokyo, Tokyo 113-8656, Japan}
\affiliation{Advanced Institute for Materials Research, Tohoku University, Sendai 980-8577, Japan}

\author{Eiji Saitoh}
\affiliation{Department of Applied Physics, Faculty of Engineering, University of Tokyo, Tokyo 113-8656, Japan}
\affiliation{Institute for AI and Beyond, The University of Tokyo, Tokyo 113-8656, Japan}
\affiliation{Quantum-Phase Electronics Center, University of Tokyo, Tokyo 113-8656, Japan}
\affiliation{Advanced Institute for Materials Research, Tohoku University, Sendai 980-8577, Japan}
\affiliation{Advanced Science Research Center, Japan Atomic Energy Agency, Tokai 319-1195, Japan.}

\date{\today}

\begin{abstract}
Parametric oscillation occurs when \textcolor{black}{the resonance frequency} of an oscillator is periodically modulated. Owing to time-reversal symmetry breaking in magnets, nonreciprocal magnons can be parametrically excited when spatial-inversion symmetry breaking is provided. This means that magnons with opposite propagation directions have different amplitudes. Here we demonstrate switching on and off the magnon parametric oscillation by reversing the external field direction applied to a Y$_3$Fe$_5$O$_{12}$ micro-structured film. The result originates from the nonreciprocity of surface mode magnons, leading to field-direction dependence of the magnon accumulation under a nonuniform microwave pumping. Our numerical calculation well reproduces the experimental result.
\end{abstract}

\maketitle
\begin{figure}
\includegraphics[width=\columnwidth]{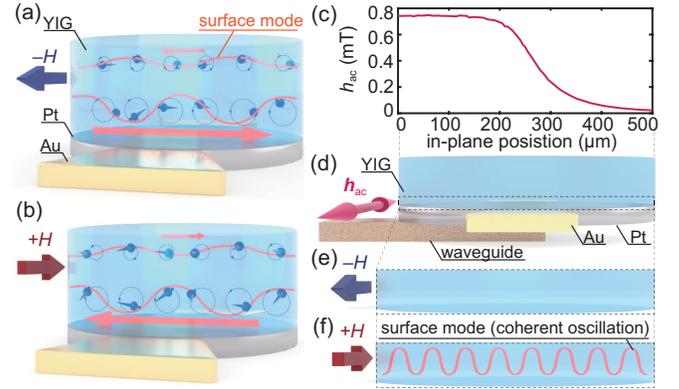}
\caption{(a), (b) Schematic of the fabricated sample and the surface mode magnons. The nonreciprocal surface mode magnons propagate oppositely for opposite field directions. The bottom surface has larger surface mode amplitude because it is closer to a CPW. (c) A calculation result of in-plane nonuniform microwave field distribution, simulated on COMSOL. Simulation was performed with the same waveguide dimensions as in the experiment, setting the input microwave power to 300 mW. (d) Schematic of the sample and the CPW. The sample is partially sticking out of the CPW, leading to nonuniform excitation as in (c). (e), (f) Concept of field-direction dependent parametric excitation. Here, surface mode magnons are not excited when the external field is in the $-H$ direction while they coherently oscillate for $+H$.}
\label{fig:f1}
\end{figure}

Nonreciprocity is a fundamental and universal concept in physics where opposite-direction components of a phenomenon have different properties. It emerges in symmetry-broken systems\cite{SSeki}, including those breaking time-reversal and spatial-inversion symmetries simultaneously\cite{GGitgeatpong}. In magnetic materials with such symmetry, their magnetic response can be altered when an external magnetic field is reversed. Magnon parametric excitation, where a $2f$-frequency microwave magnetic field excites spin-wave quanta called magnons with $1f$\cite{FGuo,TBracher,SMRezende,TMakiuchi,THioki_tomo,HShimizu,KHoshi,THioki}, is one type of magnetic response. In general, parametric excitation is a nonlinear process where a $2f$ modulation is given to a parameter of an oscillator to excite a $1f$ oscillation\cite{LDLandau}, and its applications are not limited to magnetic materials (magnon parametrons)\cite{TMakiuchi,THioki_tomo,HShimizu,KHoshi,THioki} but also optical parametric oscillators\cite{JAGiordmaine} or Josephson parametric amplifiers\cite{BYurke}. Importantly, magnon parametrons break time-reversal symmetry due to their magnetization. This allows surface (Damon-Eshbach) mode magnons, which break inversion symmetry, to have nonreciprocity\cite{SMRezende,RWDamon,MJHurben,DDStancil,TAn,KYamamoto}. Therefore, their propagation direction is determined by external field direction and can be reversed by field direction reversal [Figs. \ref{fig:f1}(a) and (b)]. This property leads to a fascinating hypothesis: Field-direction dependent parametric oscillation could be achieved by parametrically exciting the surface mode magnons.
 Here, we demonstrate switching on and off the parametric oscillation by reversing the external field direction [Figs. \ref{fig:f1}(e) and (f)]. We used an yttrium iron garnet (Y$_3$Fe$_5$O$_{12}$; YIG) disk. YIG is a model material for magnonics being a ferrimagnet with low magnetocrystalline anisotropy and magnetic damping\cite{SMRezende}.

\begin{figure*}
\includegraphics[width=\textwidth]{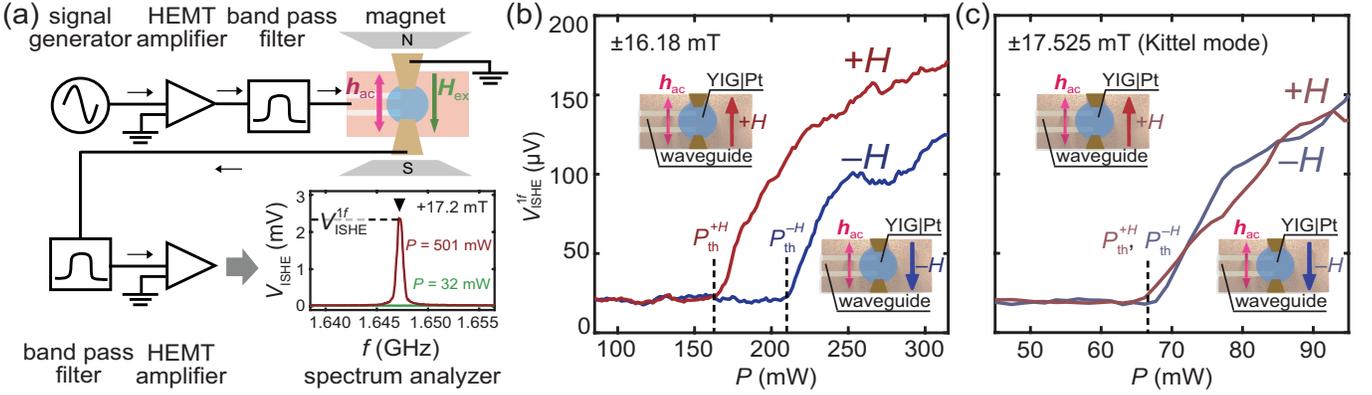}
\caption{(a) Schematic of the measurement setup. The green curve (32 mW) in the inset shows that the peak at $1f$ did not emerge below the parametric oscillation threshold. In contrast, the red curve (501 mW) in the inset shows that the peak emerged over the threshold. (b) Power dependence of $V^{1f}_\mathrm{ISHE}$ for a fixed field magnitude 16.18 mT. (c) Power dependence of $V^{1f}_\mathrm{ISHE}$ for a fixed field magnitude 17.525 mT (\textcolor{black}{Kittel mode excitation}).}
\label{fig:f2}
\end{figure*}
In this study, we sputtered a Pt(5 nm) film on top of YIG(1.4 \si{\micro\metre}) grown on a gadolinium gallium garnet (Gd$_3$Ga$_5$O$_{12}$; GGG) substrate by liquid phase epitaxy. We fabricated the film into a 500-\si{\micro\metre}-diameter disk using a negative photolithography process. Two gold electrodes were then attached onto the disk using a positive photolithography process. The Pt side of the fabricated sample was mounted onto a microwave coplanar waveguide (CPW). The sample partially stuck out of the CPW so that we can excite the sample nonuniformly in the in-plane direction, as shown in Figs. \ref{fig:f1}(c) and (d). \textcolor{black}{Here, Fig. \ref{fig:f1}(c) was obtained by a simulation on COMSOL using the same waveguide dimensions as the experiment.} Ferromagnetic resonance (FMR) in YIG from the same YIG$|$GGG batch was measured for another Pt$|$YIG disk on a microwave CPW to measure the saturation magnetization $\mu_0M_\mathrm{s} = 148.8$ mT, and for a YIG film in a microwave cavity to measure the Gilbert damping constant $\alpha\sim 2.1\times10^{-3}$, the uniaxial (out-of-plane) anisotropy $K_\mathrm{u} = –1707.6$ J m$^{-3}$ and the cubic anisotropy $K_1 = –44.4$ J m$^{-3}$.

The sample with the CPW was placed between an electromagnet in the measurement setup depicted in Fig. \ref{fig:f2}(a). The external field of the electromagnet was applied in the in-plane direction of the sample so that the surface mode magnons can be excited\cite{SMRezende,RWDamon,MJHurben,DDStancil,TAn}. A signal generator transmitted a pumping microwave with a fixed frequency of $2f = 3.294$ GHz to the sample. The signal was amplified by a 35 dB \textcolor{black}{high electron mobility transistor (HEMT)} low-noise amplifier and passed through a 2-4 GHz band-pass filter. The microwave magnetic field was applied parallel to the external field, exciting $1f = 1.647$ GHz magnons parametrically in the YIG disk. The magnons generated an AC spin current via the AC spin pumping\cite{HJJiao,CHahn,DWei,PHyde,MWeiler} at the YIG$|$Pt interface. The AC spin current then converted into an AC electric current via the inverse spin Hall effect (ISHE)\cite{YTserkovnyak,SMizukami,AAzevedo,ESaitoh,MVCostache,SOValenzuela,TKimura,YKajiwara,KAndo} in the Pt layer. The signal went through a 1-2.5 GHz band-pass filter, amplified by another 35 dB HEMT low-noise amplifier. The AC ISHE voltage $V_\mathrm{ISHE}$ between the two gold electrodes was measured by a spectrum analyzer, and the $1f$ component $V^{1f}_\mathrm{ISHE}$ was recorded. Data was taken for 50 times and the mean value was recorded. \textcolor{black}{Because the Pt layer touched the waveguide, the 2$f$ AC current for microwave generation may have entered the spectrum analyzer, but this does not account for $V^{1f}_\mathrm{ISHE}$ due to the 1$f$ frequency selection on the signal analyzer. Moreover, the 1$f$ artifact due to nonlinearity in the HEMT amplifier was removed by the 2-4 GHz band-pass filter before reaching the sample. Note that an inductive voltage due to magnetization precession in the YIG layer could account for $V^{1f}_\mathrm{ISHE}$, but this can be considered small compared to the ISHE contribution in a low microwave power regime\cite{CHahn}.}We conducted measurements sweeping the microwave power $P$ from 31.6 to 501 mW, and the external field $H_\mathrm{ex}$ from --13 to --19 mT ($-H$) or from 13 to 19 mT ($+H$). 

Increasing the power at a specific external field causes an ISHE voltage peak at $1f$ to emerge over a critical power, as shown in the plot of Fig. \ref{fig:f2}(a). Figure \ref{fig:f2}(b) shows the observed $1f$ ISHE voltage as a function of the microwave power for both the field directions ($-H$ and $+H$) with the fixed magnitude, which is tuned to excite finite-wavenumber magnons. A significant threshold difference is observed for different field directions. On the other hand, when the magnetic field is tuned to excite uniform magnetization precession (\textcolor{black}{Kittel mode}), the threshold is almost the same for both the field directions, as shown in Fig. \ref{fig:f2}(c). For Figs. \ref{fig:f2}(b) and (c), a moving average was taken for five previous datapoints.

\begin{figure}
\includegraphics[width=\columnwidth]{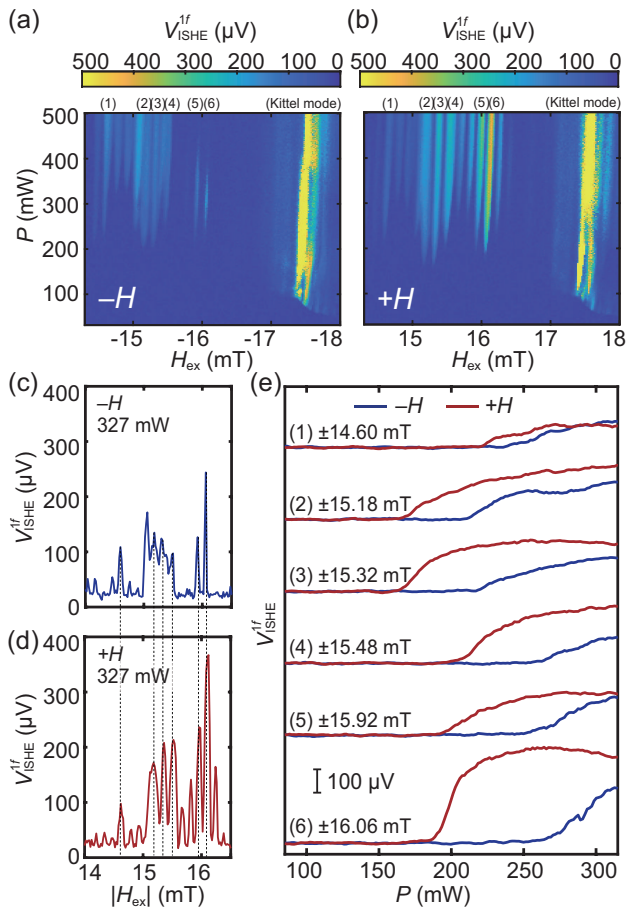}
\caption{(a), (b) Field and power dependence of the $1f$ component of the ISHE voltage $V^{1f}_\mathrm{ISHE}$ for $-H$ and $+H$ cases, showing peak structures. (c), (d) Field dependence of $V^{1f}_\mathrm{ISHE}$ obtained by fixing the power to 327 mW in (a) and (b). (e) Power dependence of \textcolor{black}{$V_\mathrm{ISHE}^{1f}$ taken at different external magnetic field conditions (1)-(6).}}
\label{fig:f3}
\end{figure}

Figures \ref{fig:f3}(a) and (b) show the $1f$ ISHE voltage when sweeping the external field value and the microwave power. Data was taken for 50 times and the mean value was recorded for each datapoint. Peak structures were observed for both field directions, where multiple low peaks (one high peak) correspond to the finite-wavenumber magnons (\textcolor{black}{Kittel mode} magnons)\cite{FGuo}. The field dependence of the ISHE voltage was obtained by fixing the microwave power to 327 mW, shown in Figs. \ref{fig:f3}(c) and (d). Here, we checked that the difference of peak field magnitudes is negligible between both field directions, allowing to compare the peaks in Figs. \ref{fig:f3}(a) and (b) at the same field magnitude. Fig. \ref{fig:f3}(e) is the power dependence plot for each pair of finite-wavenumber magnon modes, exhibiting significantly lower thresholds in all the $+H$ cases. A moving average was taken for five previous data points. The ISHE voltage ratio reached up to around 700$\%$ between the thresholds for $-H$ and $+H$.

The negligible threshold difference in Fig. \ref{fig:f2}(c) is due to the reciprocal nature of the \textcolor{black}{Kittel mode}, while the significant threshold difference in Fig. \ref{fig:f3}(e) requires the nonreciprocal surface mode magnons to be excited. However, the excitation \textcolor{black}{of the surface mode} alone cannot explain the threshold difference; nonuniform pumping microwave is necessary for both out-of-plane and in-plane directions. This is because out-of-plane (in-plane) uniform excitation would result in identical situations for opposite field directions when viewed at 180 degrees around the in-plane (out-of-plane) axis. The out-of-plane nonuniformity is provided by the sample thickness, which leads to microwave attenuation with distance [Figs. \ref{fig:f1}(a) and (b)]. The in-plane nonuniformity is achieved by partially sticking the sample out of the CPW, which leads to stronger pumping on the waveguide and weaker pumping out of it [Fig. \ref{fig:f1}(c)]. Due to the out-of-plane nonuniformity, we only consider magnon excitation on the bottom surface (the surface closer to the CPW). In the $+H$ case, thermal magnons on the bottom surface propagate towards the strongly-pumped region (the left side of Fig. \ref{fig:f1}(d)) due to the nonreciprocity of the surface mode magnons. Accumulation at the left edge follows the propagation because \textcolor{black}{the conductance of magnons between both the surfaces is suppressed by two-magnon interaction \cite{TAn, AGGurevich,AVChumak}}. Due to the strong pumping, the accumulated thermal magnons experience small effective damping on average. This makes it easy for magnons to overcome linear magnetic damping and coherently oscillate [Fig. \ref{fig:f1}(f)], leading to lower thresholds. This is represented in the formulation of parametric oscillation threshold for a $k$-wavenumber mode, 

\begin{equation}
\label{eq:e1}
h_{k,\mathrm{crit}}=\frac{2\pi\sqrt{(f_k\alpha_k)^2+\Delta f_k^2}}{|\rho_k|},
\end{equation}
where $f_k$ is the frequency of the excited mode, $\alpha_k$ is the linear damping constant, $\Delta f_k = f_k - f$ is the detuning term, and $\rho_k$ is the coupling constant between a photon and magnons, determined by material parameters of YIG\cite{TBracher,SMRezende}. In contrast, in the $-H$ case, thermal magnons propagate towards the weakly-pumped region (the right side of Fig. \ref{fig:f1}(d)) and accumulate in the right edge. Accordingly, the thermal magnons experience large effective damping on average, which makes it difficult for magnon to coherently oscillate [Fig. \ref{fig:f1}(e)], leading to higher thresholds. Since the microwave decay length is longer for out-of-plane than for in-plane, the out-of-plane decay does not give much effective damping to magnons in the $+H$ case. Nonlinear magnon interactions\cite{SMRezende} could affect the threshold by altering the detuning term $\Delta f_k$ in Eq. (\ref{eq:e1}), but this is not significant due to the small change in the detuning term, approximated by multiplying the difference of the peak field magnitude ($\sim0.02$ mT) in Figs. \ref{fig:f3}(c) and (d) by the gyromagnetic ratio $\gamma = 2\pi\times2.8025\times10^{10}$ Hz T$^{-1}$. Note that the fine peak structures in Figs. \ref{fig:f3}(a) and (b) are explained by the interference between oppositely propagating surface mode magnons on both the surfaces, generating odd or even standing wave modes in the sample volume\cite{FGuo}.

\begin{figure}
\includegraphics[width=\columnwidth]{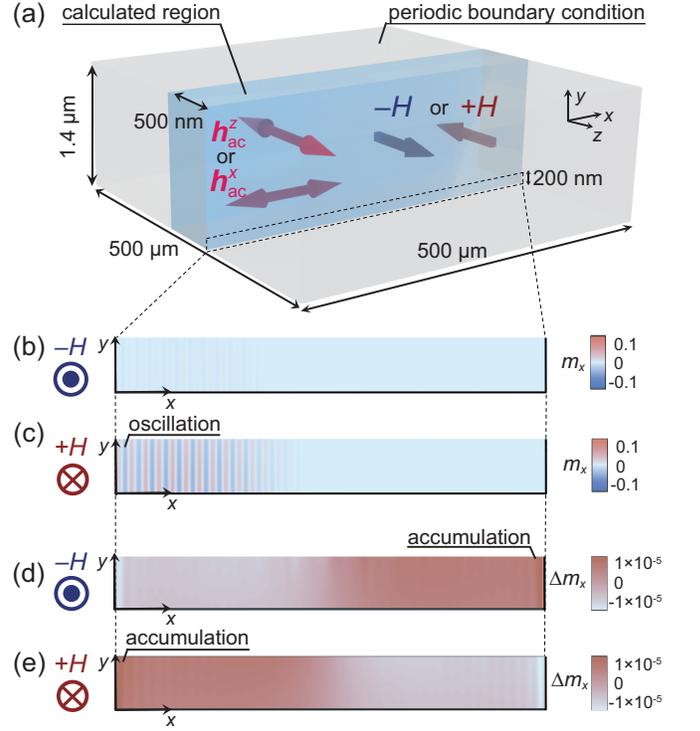}
\caption{(a) Schematic of numerical calculation setup. Calculation was performed on a YIG cuboid thin film (\textcolor{black}{500 \si{\micro\metre}$\times$1.4 \si{\micro\metre}$\times$500 nm}), the $y$-direction being out of plane. We used a periodic boundary condition of \textcolor{black}{500 \si{\micro\metre}} in the $z$-direction. (b), (c) The calculation result of $m_x$ (the $x$-component of the magnetization) during the parametric excitation, zoomed into the bottom \textcolor{black}{200 nm} of (a), normalized by the saturation magnetization.  \textcolor{black}{(d), (e) The calculation result of $\Delta m_x$ (the in-plane asymmetry) during the FMR, zoomed into the bottom \textcolor{black}{200 nm} of (a).} }
\label{fig:f4}
\end{figure}

We demonstrate the aforementioned experimental result and discussion by numerical calculation. The dynamics of magnetization was calculated by using the mumax3 code\cite{AVansteenkiste}, which solves the Landau--Lifshitz--Gilbert (LLG) equation. The saturation magnetization and the Gilbert damping constant were set to $\mu_0M_\mathrm{s} = \textcolor{black}{148.8}$ mT, $\alpha = \textcolor{black}{2.1\times10^{-3}}$, respectively. The effective field included the external, exchange (given by the stiffness constant $A_\mathrm{ex} = 4.73$ pJ m$^{-1}$) and demagnetizing fields. Calculation was performed on a YIG cuboid thin film (\textcolor{black}{500 \si{\micro\metre}$\times$1.4 \si{\micro\metre}$\times$500 nm}) [Fig. \ref{fig:f4}(a)], where the $y$-direction was out of plane. A periodic boundary condition of \textcolor{black}{500 \si{\micro\metre}} was imposed in the $z$-direction, and a unit cell was set to 100 nm$\times$\textcolor{black}{10 nm$\times$500 nm}. The temperature was set to 0 K. \textcolor{black}{First, we performed a calculation to confirm whether the field-direction dependence of magnon parametric excitation reproduces the experimental result. To this end, microwave and external fields were applied in the $z$-direction.} The microwave field distribution was set to a sigmoid for the $x$- (in-plane) \textcolor{black}{direction} and an exponent for the $y$- (out-of-plane) \textcolor{black}{direction}, represented by
\begin{equation}
\label{eq:e2}
 h_\mathrm{ac}^z(x, y, t) = h_0^z[1 + \exp(x/x_0)]^{-1}\exp(-y/\lambda)\sin(2\pi\times2ft), 
\end{equation}
where \textcolor{black}{$h_0^z = 318$  \si{\micro\tesla}, $1f = 1.66533$ GHz, $x_0 =16$ \si{\micro\metre} and  $\lambda =456$ \si{\micro\metre}. The spatial profile $x_0$ and $\lambda$ of the microwave was estimated from the COMSOL simulation performed for the same parameters as those for Fig. \ref{fig:f1}(c)}. The external field magnitude was set to \textcolor{black}{16} mT ($<$ \textcolor{black}{Kittel mode} field \textcolor{black}{17.5} mT) to excite finite-wavenumber magnons. The external field direction was set to either $-H$ or $+H$, respectively corresponding to the $+z$ or $-z$ direction in Fig. \ref{fig:f4}(a). \textcolor{black}{Magnon oscillation is barely observed in the $-H$ case [Fig. \ref{fig:f4}(b)] while it is evident in the $+H$ case [Fig. \ref{fig:f4}(c)]. }

\textcolor{black}{Next, we performed a calculation to confirm the appearance of the magnon accumulation in the film. To this end, the microwave and external fields were applied in the $x$- and $z$-directions, respectively, to excite FMR. The microwave field distribution was set uniform for the $x$- (in-plane) direction and to an exponent for the $y$- (out-of-plane) direction, represented by
\begin{equation}
\label{eq:e3}
 h_\mathrm{ac}^x(y, t) = h_0^x\exp(-y/\lambda)\sin(2\pi\times1ft), 
\end{equation}
where $h_0^x = 10$ \si{\micro\tesla}, and the other values were the same to those in the parametric excitation case. The amplitude envelope was extracted from the calculation result and large-wavenumber components due to calculation error were omitted. After these procedures, we obtained the spatial distribution of in-plane asymmetry $\Delta m_x$ by subtracting the $m_x$ distribution flipped over in the $x$-direction from the original distribution. From the obtained $\Delta m_x$ distribution, we confirmed that the accumulation is larger in the right (left) edge of the bottom surface in $-H$ ($+H$) case as shown in Fig. \ref{fig:f4}(d) [Fig. \ref{fig:f4}(e)], corresponding to the discussion.}
 
In summary, we found that the thresholds in the magnon parametron significantly \textcolor{black}{change} by reversing the external field direction. This results from the accumulation of the nonreciprocal surface mode magnons under the nonuniform microwave field, which can be easily achieved in experiments. We also obtained the numerical calculation well reproducing the experimental result \textcolor{black}{and the discussion}. The threshold changes observed in this study enables the switching behavior \textcolor{black}{at a microwave power} between the thresholds for both the field directions, which can be applied to making logic elements. The high switchability of the magnetization dynamics in terms of field direction will present great advantages in designing magnon-based computing devices. Furthermore, it paves an experimental way for the crossover of nonlinear phenomena and nonreciprocity in magnetic materials.

\begin{acknowledgments}
This work was supported by JST CREST (JPMJCR20C1 and JPMJCR20T2),  JSPS KAKENHI (JP19H05600,  JP20K15160, and JP22K14584), Institute for AI and Beyond of the University of Tokyo, IBM–UTokyo lab, and Advanced Technology Institute Research Grants 2022.
\end{acknowledgments}

\clearpage

\end{document}